\documentclass{article}

\usepackage{cite}
\usepackage{graphicx}
\usepackage{dcolumn}

\begin{document}

\date{}
\title{On a model for rotational tunneling with a $C_{6}$-space-time symetric analog}
\author{Paolo Amore \\
Facultad de Ciencias, CUICBAS, Universidad de Colima, \\
Bernal D\'{\i}az del Castillo 340, Colima, Colima, Mexico \thanks{%
paolo.amore@gmail.com}\\
Francisco M. Fern\'andez \\
INIFTA (CONICET), Divisi\'{o}n Qu\'{i}mica Te\'{o}rica, \\
Blvd. 113 y 64 (S/N), Sucursal 4, Casilla de Correo 16, \\
1900 La Plata, Argentina \thanks{%
framfer@gmail.com}}
\maketitle

\begin{abstract}
We analyze the simple model of a rigid rotor with $C_{3}$ symmetry and show
that the use of parity simplifies considerably the calculation of its
eigenvalues. We also consider a non-Hermitian space-time-symmetric
counterpart that exhibits real eigenvalues and determine the exceptional
point at which the antiunitary symmetry is broken.
\end{abstract}

\section{Introduction}

\label{sec:intro}

Rotational tunnelling takes place when groups of atoms in a
molecule rotate, as an almost rigid structure, about a single
bond. When the barriers between different nuclear configurations
are sufficiently high some of the lowest states exhibit close
energies and the transition between them can be investigated by
suitable spectroscopies\cite{CHHP84,PH97,H99}. A typical example
is provided by the methyl group ($-CH_{3}$).

Rotational symmetry is commonly studied by means of simple models based on
effective Hamiltonians for properly chosen restricted rigid rotors\cite
{CHHP84,PH97,H99,HH85,KS16,KS17}. In some cases a single rotor model
provides an acceptable description of the experimental data but in others
one has to resort to a set of coupled rotors. The Schr\"{o}dinger equation
for such models has been solved in more than one way\cite{HH85,KS17}.

In this paper we are interested in a well known algorithm for the solution
of band matrices\cite{Z83,Z84, ZST85,FOT86} that may be a convenient
alternative approach to the iterative matrix inversion proposed several
years ago\cite{HH85}. Although today the diagonalization of a band matrix
offers no difficulty we think that such alternative methods may still be of
interest. In addition to the comparison of the methods for solving the
eigenvalue equation we want to point out that the case of a small rotational
barrier (or large quantum numbers) may lead to numerical errors due to
almost degenerate rotational states.

In addition to what has just mentioned we will also discuss a space-time ($%
\mathcal{ST}$) symmetric non-Hermitian version of the effective Hamiltonian
for the restricted rotor that takes place when the barrier height is allowed
to be purely imaginary. This kind of problems have been intensely studied in
recent years (see\cite{B07} for an earlier review on the issue and also\cite
{BK11,FG14} for closely related models).

In section~\ref{sec:model} we discuss the problem of nearly degenerate
energies by means of a simple rigid-rotor model with symmetry $C_{3}$. In
section~\ref{sec:parity} we show how to go around such difficulty by means
of symmetry arguments. In section~\ref{sec:S-T_symmetry} we consider the $%
\mathcal{ST}$-symmetric non-Hermitian counterpart and determine the regions
of exact and broken $\mathcal{ST}$ symmetry. Finally, in section~\ref
{sec:conclusions} we summarize the main results of the paper and draw
conclusions.

\section{Restricted-rotor model}

\label{sec:model}

For concreteness, in this paper we consider the rotation of a group of atoms
hindered by a potential $V(\phi )$ with periodicity $V(\phi +2\pi /3)=V(\phi
)$. It is commonly expanded in a Fourier series of the form\cite
{CHHP84,PH97,H99}
\begin{equation}
V(\phi )=\sum_{j=0}^{\infty }V_{3j}\cos (3j\phi ).  \label{eq:V_Fourier}
\end{equation}
For present discussion it is sufficient to consider just the leading term so
that the hindered rotator is given by the effective Hamiltonian operator
\begin{equation}
H=-B\frac{d^{2}}{d\phi ^{2}}+V(\phi ),\;V(\phi )=V_{3}\cos \left( 3\phi
\right) ,  \label{eq:H}
\end{equation}
where the magnitude of the rotational constant $B=\hbar ^{2}/(2I)$ is
determined by the moment of inertia $I$ of the rotor. It is convenient to
measure the energy $E$ in units of $B$ so that the dimensionless
Schr\"{o}dinger equation becomes
\begin{eqnarray}
H\psi &=&\epsilon \psi ,  \nonumber \\
H &=&-\frac{d^{2}}{d\phi ^{2}}+V(\phi ),\;V(\phi )=\lambda \cos \left( 3\phi
\right) ,  \nonumber \\
\epsilon &=&\frac{E}{B},\;\lambda =\frac{V_{3}}{B}.  \label{eq:Schro_dim}
\end{eqnarray}

Since the potential is periodic of period $2\pi /3$ the eigenfunctions form
basis for the irreducible representations $A$ and $E$ of the symmetry group $%
C_{3}$. Therefore, the Fourier expansions for the eigenfunctions are of the
form
\begin{equation}
\psi _{s}(\phi )=\sum_{j=-\infty }^{\infty }c_{j,s}f_{j,s}(\phi
),\;f_{j,s}(\phi )=\frac{1}{\sqrt{2\pi }}e^{i\left( 3j\phi +s\right)
},\;s=0,\pm 1,  \label{eq:Fourier_exp}
\end{equation}
where the subscripts $s=0$ and $s=\pm 1$ correspond to the symmetry species $%
A$ and $E$, respectively. By means of the Fourier expansions (\ref
{eq:Fourier_exp}) the Schr\"{o}dinger equation (\ref{eq:Schro_dim}) becomes
a three-diagonal secular equation
\begin{equation}
\lambda c_{m-1,s}+2\left[ \epsilon -\left( 3m+s\right) ^{2}\right]
c_{m,s}+\lambda c_{m+1,s}=0,\;m=0,\pm 1,\pm 2,\ldots .
\label{eq:recurrence_relation}
\end{equation}

In practice we truncate the secular equation
(\ref{eq:recurrence_relation}) and solve a matrix eigenvalue
problem of dimension, say, $2N+1$. However, some time ago
H\"{a}usler and H\"{u}ller\cite{HH85} proposed an iterative
method, based on matrix inversion, that avoids matrix
diagonalization. Today, such diagonalization can be carried out
most easily even in the most modest personal computer.
Nonetheless, we want to point out to an even simpler strategy
proposed some time ago\cite{Z83,Z84, ZST85,FOT86} that consists in
solving the secular equation (\ref{eq:recurrence_relation}) as a
recurrence relation. The truncation of the secular equation just
mentioned is equivalent to setting the boundary conditions
$c_{m,s}=0$ for $|m|>N$ in the
recurrence relation (\ref{eq:recurrence_relation}). Therefore, if we set $%
c_{-N,s}=1$ we can calculate $c_{j,s}$ for $j=-N+1,-N+2,\ldots $ so that the
roots of $c_{N+1,s}(\epsilon )=0$ are exactly the roots of the
characteristic polynomial of the secular matrix of dimension $2N+1$ that
yield estimates of the energies of the problem.

In what follows $\epsilon _{0,s}(\lambda )<\epsilon _{1,s}(\lambda
)<\epsilon _{2,s}(\lambda )<\ldots $denote the energies of the hindered
rotor. When $\lambda =0$ the $A$ states are $\epsilon _{0,0}^{(0)}=0$, $%
\epsilon _{2n-1,0}^{(0)}=\epsilon _{2n,0}^{(0)}=9n^{2}$, $n=1,2,\ldots $. On
the other hand, the $E$ states are doubly degenerate for all $\lambda \geq 0$
and for $\lambda =0$ satisfy $(3n-1)^{2}=(-3n+1)^{2}$ which are obviously
treated separately. In other words, the hindered potential splits the doubly
degenerate $A$ states while the $E$ ones can be treated as nondegenerate
with symmetry quantum numbers $s=-1$ ($E_{a}$) and $s=1$ ($E_{b}$). For this
reason the calculation of the latter eigenvalues is much simpler.

\begin{figure}[]
\begin{center}
\includegraphics[width=9cm]{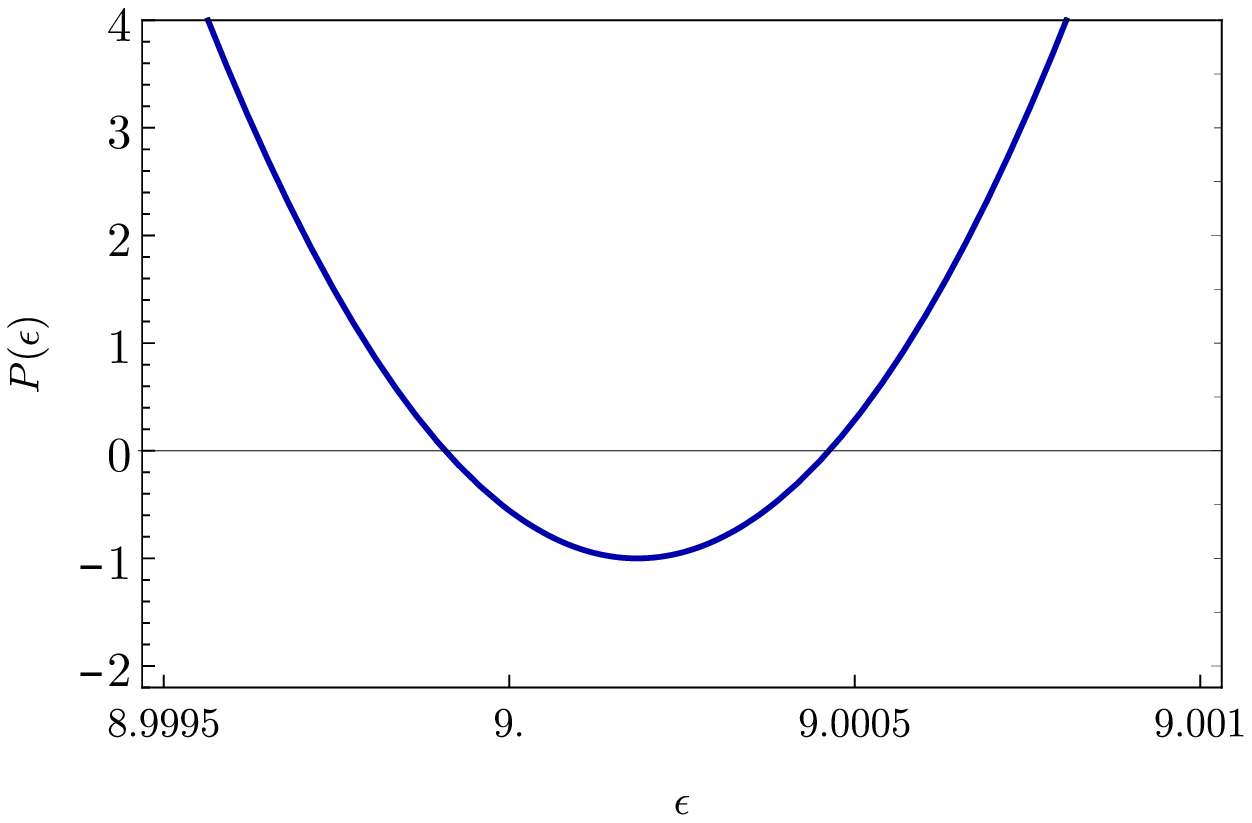} %
\includegraphics[width=9cm]{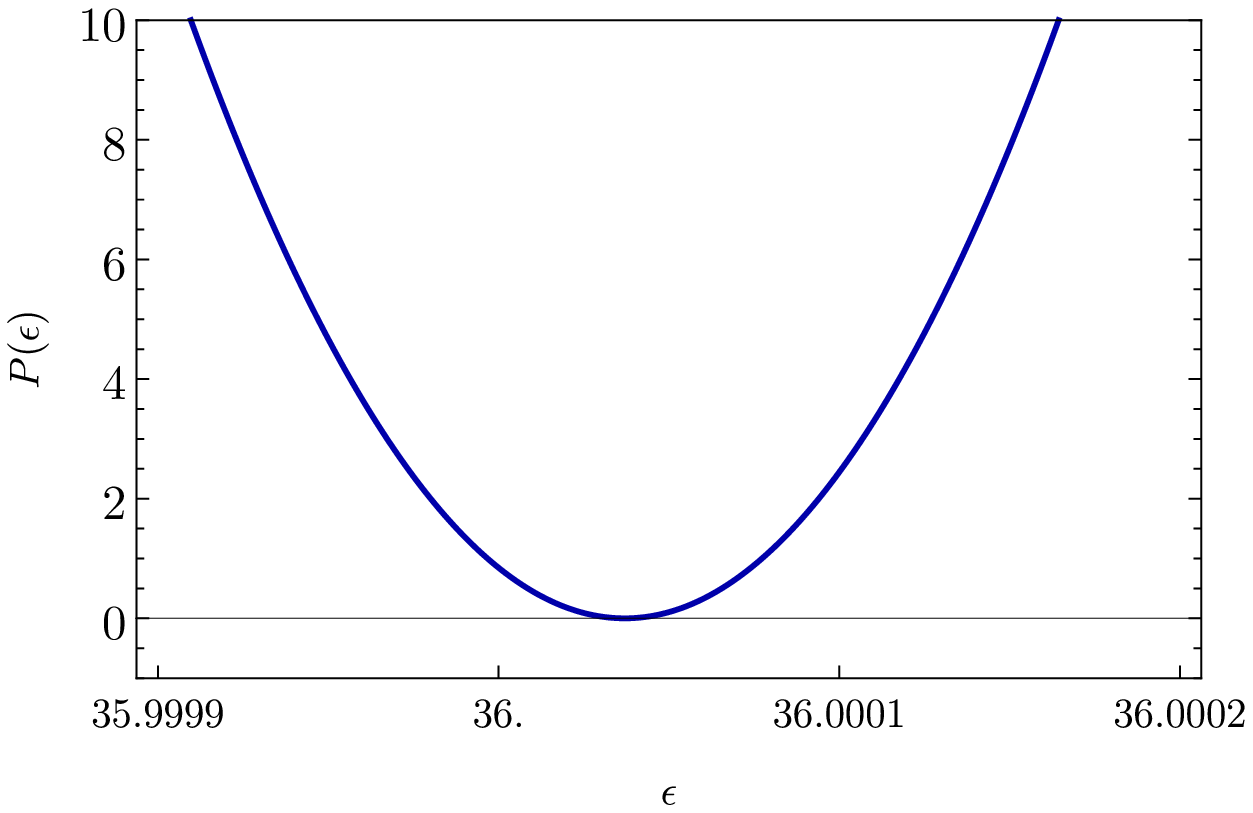}
\end{center}
\caption{Scaled characteristic polynomial $P(\epsilon)$ for $\lambda=0.1$}
\label{fig:P(epsilon)}
\end{figure}

When $\lambda $ is sufficiently small the eigenvalues $\epsilon
_{2n-1,0}(\lambda )$ and $\epsilon _{2n,0}(\lambda )$ are quasi degenerate
which may make their numerical calculation somewhat difficult. An example is
given in Figure~\ref{fig:P(epsilon)} that shows the characteristic
polynomial $P(\epsilon )$ for $\lambda =0.1$ properly scaled to reduce its
size. We clearly see that the splitting of the degenerate states is
considerably smaller for $n=2$ than for $n=1$. In general, the magnitude of
the splitting $\epsilon _{2n,0}(\lambda )-\epsilon _{2n-1,0}(\lambda )$
decreases as $n$ increases so that the problem also appears for greater
values of $\lambda $ if the quantum number is large enough. Some algorithms
may fail to find the almost identical roots of $P(\epsilon )$ if the
accuracy of the calculation is insufficient. For $\lambda =0.1$ the
corresponding eigenvalues are $\epsilon _{1,0}=8.99990740760586$, $\epsilon
_{2,0}=9.00046293268167$, $\epsilon _{3,0}=36.0000370368357$ and $\epsilon
_{4,0}=36.0000370373120$.

The application of perturbation theory is most revealing. When $\lambda \neq
0$ the perturbation expansions for the first $A$ eigenvalues are
\begin{eqnarray}
\epsilon _{0,0} &=&-\frac{1}{18}\lambda ^{2}+\frac{7}{23328}\lambda ^{4}-%
\frac{29}{8503056}\lambda ^{6}+\ldots ,  \nonumber \\
\epsilon _{1,0} &=&9-\frac{1}{108}\lambda ^{2}+\frac{5}{2519424}\lambda ^{4}-%
\frac{289}{293865615360}\lambda ^{6}+\ldots ,  \nonumber \\
\epsilon _{2,0} &=&9+\frac{5}{108}\lambda ^{2}-\frac{763}{2519424}\lambda
^{4}+\frac{1002401}{293865615360}\lambda ^{6}+\ldots ,  \nonumber \\
\epsilon _{3,0} &=&36+\frac{1}{270}\lambda ^{2}-\frac{317}{157464000}\lambda
^{4}+\frac{10049}{10044234900000}\lambda ^{6}+\ldots ,  \nonumber \\
\epsilon _{4,0} &=&36+\frac{1}{270}\lambda ^{2}+\frac{433}{157464000}\lambda
^{4}-\frac{5701}{10044234900000}\lambda ^{6}+\ldots ,  \nonumber \\
\epsilon _{5,0} &=&81+\frac{1}{630}\lambda ^{2}+\frac{187}{8001504000}%
\lambda ^{4}-\frac{5861633}{342986069260800000}\lambda ^{6}+\ldots ,
\nonumber \\
\epsilon _{6,0} &=&81+\frac{1}{630}\lambda ^{2}+\frac{187}{8001504000}%
\lambda ^{4}+\frac{6743617}{342986069260800000}\lambda ^{6}+\ldots .
\label{eq:PT_series_A}
\end{eqnarray}
We appreciate that $\epsilon _{2n,0}(\lambda )-\epsilon _{2n-1,0}(\lambda
)=O(\lambda ^{2n})$. We did not apply the standard perturbation theory for
degenerate states\cite{M76} because it is rather impractical in the present
case; instead we obtained the perturbation expansions (\ref{eq:PT_series_A})
from the characteristic polynomial for sufficiently large values of $N$.

\section{Parity}

\label{sec:parity}

In order to solve the problem posed by the quasi-degenerate $A$ states we
take into account that the potential is parity invariant: $V(-\phi )=V(\phi )
$. If $P$ denotes the parity operator then $P\psi _{s}(\phi )=\psi
_{s}(-\phi )=\psi _{-s}(\phi )$ transforms states $E_{a}$ into $E_{b}$ but
the $A$ states remain as such. This fact allows us to separate the latter
states into even and odd ones:
\begin{eqnarray}
\psi _{A_{+}}(\phi ) &=&c_{0}\frac{1}{\sqrt{2\pi }}+\sum_{j=1}^{\infty }c_{j}%
\frac{1}{\sqrt{\pi }}\cos (3j\phi ),  \nonumber \\
\psi _{A_{-}}(\phi ) &=&\sum_{j=1}^{\infty }c_{j}\frac{1}{\sqrt{\pi }}\sin
(3j\phi ).  \label{eq:psi_A+-}
\end{eqnarray}
In this way we have a secular equation
\begin{eqnarray}
\epsilon c_{0}+\frac{\lambda }{\sqrt{2}}c_{1} &=&0,  \nonumber \\
\frac{\lambda }{\sqrt{2}}c_{0}+(\epsilon -9)c_{1}+\frac{\lambda }{2}c_{2}
&=&0,  \nonumber \\
\frac{\lambda }{2}c_{n-1}+\left( \epsilon -9n^{2}\right) c_{n}+\frac{\lambda
}{2}c_{n+1} &=&0,\;n=2,3,\ldots ,  \label{eq:rec_rel_A+}
\end{eqnarray}
for the $A_{+}$ states and another one
\begin{equation}
\frac{\lambda }{2}c_{n-1}+\left( \epsilon -9n^{2}\right) c_{n}+\frac{\lambda
}{2}c_{n+1}=0,\;n=2,3,\ldots ,  \label{eq:rec_rel_A-}
\end{equation}
for the $A_{-}$ states. This analysis based on parity is similar
to using the symmetry point group $C_{3v}$ where the states
labelled here as $A_{+}$ and $A_{-}$ belong to the symmetry
species $A_{1}$ and $A_{2}$, respectively, and the effect of the
parity operator is produced by one of the reflection planes
$\sigma _{v}$\cite{C90}.

In this way, the recurrence relations (or the corresponding tri-diagonal
matrices) do not exhibit close roots for any value of $\lambda $ and the
calculation is considerably simpler. If we choose $c_{j}=0$ for $j<0$ and $%
c_{0}=1$ we can calculate $c_{j}$ for all $j>0$ and obtain the $A_{+}$
eigenvalues from the termination condition $c_{N}(\epsilon )=0$ for
sufficiently large $N$. Exactly in the same way with $c_{j}=0$ for $j<1$ and
$c_{1}=1$ we obtain the $A_{-}$ energies of the restricted rotor. The
perturbation expansions for the first eigenvalues suggest that $\epsilon
_{2n-1,0}$ is $A_{+}$ while $\epsilon _{2n,0}$ is $A_{-}$.

\begin{figure}[]
\begin{center}
\includegraphics[width=9cm]{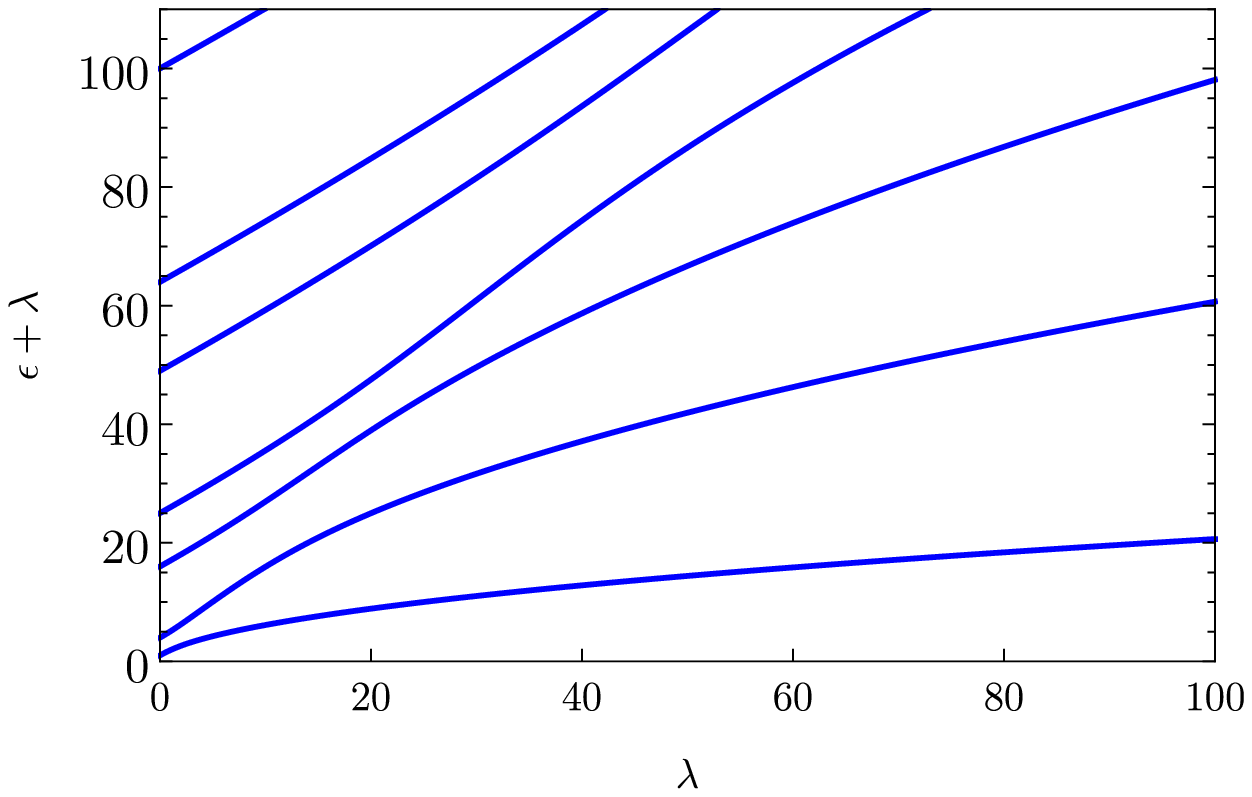}
\par
\includegraphics[width=9cm]{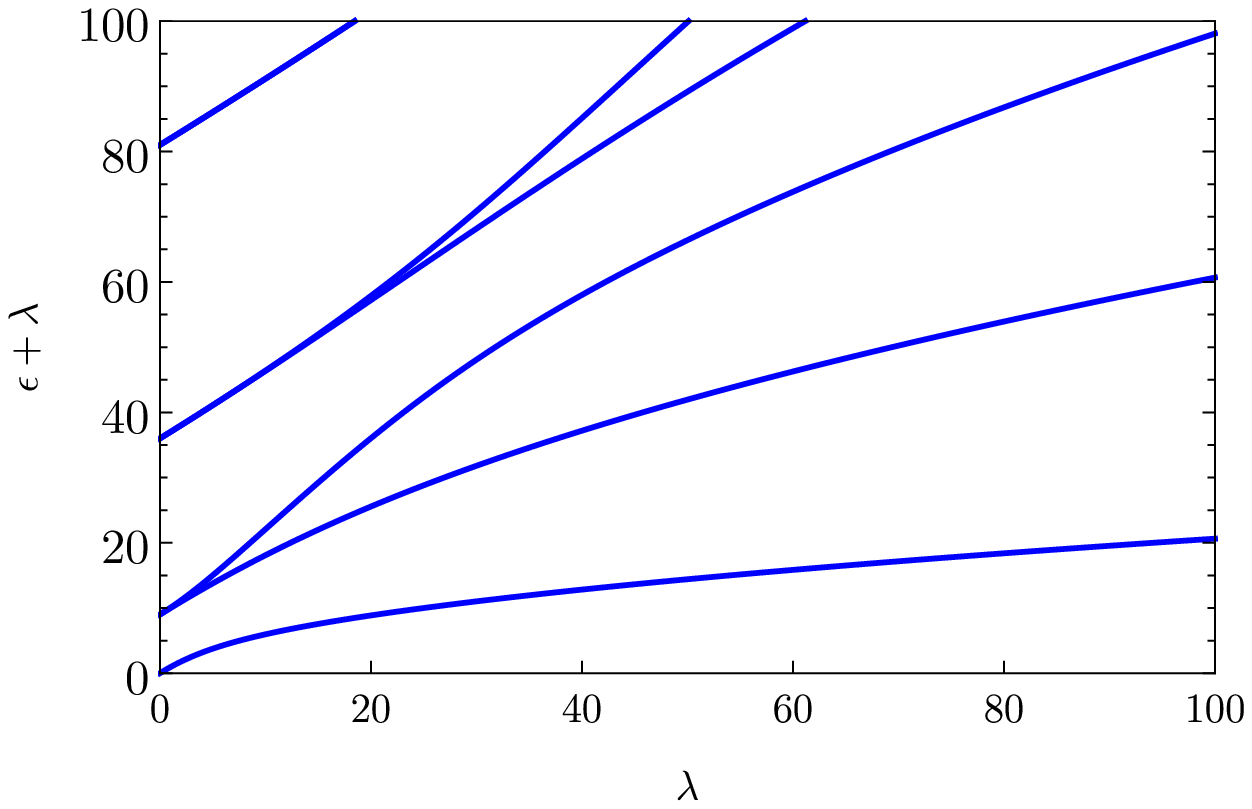}
\end{center}
\caption{Lowest eigenvalues $\epsilon(\lambda)+\lambda$ of symmetry $E$
(upper panel) and $A$ (lower panel)}
\label{fig:EPS_A_E}
\end{figure}

For large values of $\lambda $ the eigenvalues behave asymptotically as
\begin{equation}
\epsilon _{v}=-\lambda +3\sqrt{\frac{\lambda }{2}}(2v+1)+\mathcal{O}(1).
\label{eq:E_asymptotic}
\end{equation}
Figure~\ref{fig:EPS_A_E} shows the lowest eigenvalues for states of symmetry
$A$ and $E$ calculated with the expressions indicated above.

\section{Space-time symmetry}

\label{sec:S-T_symmetry}

The unitary operator $U=C_{6}$ that produces a rotation by an angle of $2\pi
/6$\cite{C90} leads to the transformation $UV(\phi )U^{-1}=V(\phi +\pi
/3)=-V(\phi )$ and $UH(\lambda )U^{-1}=H(-\lambda )$. From its application
to the eigenvalue equation $H(\lambda )\psi _{n}=\epsilon _{n}(\lambda )\psi
_{n}$, $UH(\lambda )U^{-1}U\psi _{n}=H(-\lambda )U\psi _{n}=\epsilon
_{n}(\lambda )U\psi _{n}$, we conclude that $\epsilon _{n}(\lambda )$ is
also an eigenvalue $\epsilon _{m}(-\lambda )$ of $H(-\lambda )$. Since $\psi
_{n}$ and $U\psi _{n}$ belong to the same symmetry species ($A_{+}$, $A_{-}$%
, $E_{a}$, $E_{b}$) and $\lim\limits_{\lambda \rightarrow 0}\epsilon
_{m}(-\lambda )=\lim\limits_{\lambda \rightarrow 0}\epsilon _{n}(\lambda )$
then we conclude that $m=n$ and $\epsilon _{n}(-\lambda )=\epsilon
_{n}(\lambda )$ which explains why the perturbation expansions for the
eigenvalues of $H(\lambda )$ have only even powers of $\lambda $:
\begin{equation}
\epsilon _{n}(\lambda )=\sum_{j=0}^{\infty }\epsilon _{n}^{(2j)}\lambda
^{2j}.  \label{eq:epsilon_lambda_series}
\end{equation}
This result suggests that $\epsilon _{n}(ig)$ is real for $g$
real, at least for sufficiently small values of $|g|$. This
conclusion is consistent with the fact that $H(ig)$ is
$\mathcal{ST}$ symmetric\cite{KC08,F15} with respect to the
transformation given by the antiunitary operator\cite{W60} $UT$ as
follows from $UTH(ig)TU^{-1}=H(ig)$, where $T$ is the
time-reversal operator $THT=H^{*}$ and the asterisk denotes
complex conjugation. The antiunitary symmetry tells us that the
eigenvalues are either real or appear as pairs of complex
conjugate numbers. If the antiunitary symmetry is exact ($A\psi
=a\psi $) then the eigenvalues are real, otherwise we say that it
is broken. In the present case we know, from the analysis based on
perturbation theory, that this symmetry is exact for sufficiently
small values of $|g|$.

A straightforward calculation, like the one in the preceding section,
confirms that the perturbation series for the $E$ states also have only even
powers of $\lambda $%
\begin{eqnarray}
\epsilon _{0,\pm 1} &=&1-\frac{1}{10}\lambda ^{2}+\frac{83}{32000}\lambda
^{4}-\frac{4581}{30800000}\lambda ^{6}+\ldots ,  \nonumber \\
\epsilon _{1,\pm 1} &=&4+\frac{1}{14}\lambda ^{2}-\frac{143}{54880}\lambda
^{4}+\frac{2601}{17479280}\lambda ^{6}+\ldots ,  \nonumber \\
\epsilon _{2,\pm 1} &=&16+\frac{1}{110}\lambda ^{2}+\frac{383}{37268000}%
\lambda ^{4}-\frac{72621}{958253450000}\lambda ^{6}+\ldots ,  \nonumber \\
\epsilon _{3,\pm 1} &=&25+\frac{1}{182}\lambda ^{2}+\frac{563}{385828352}%
\lambda ^{4}+\frac{144549}{30352923537664}\lambda ^{6}+\ldots ,  \nonumber \\
\epsilon _{4,\pm 1} &=&49+\frac{1}{374}\lambda ^{2}+\frac{1043}{8370179840}%
\lambda ^{4}+\frac{90081}{3366013416487040}\lambda ^{6}+\ldots .
\label{eq:PT_series_E}
\end{eqnarray}
In this case the interaction potential does not break the two-fold
degeneracy.

\begin{figure}[]
\begin{center}
\includegraphics[width=9cm]{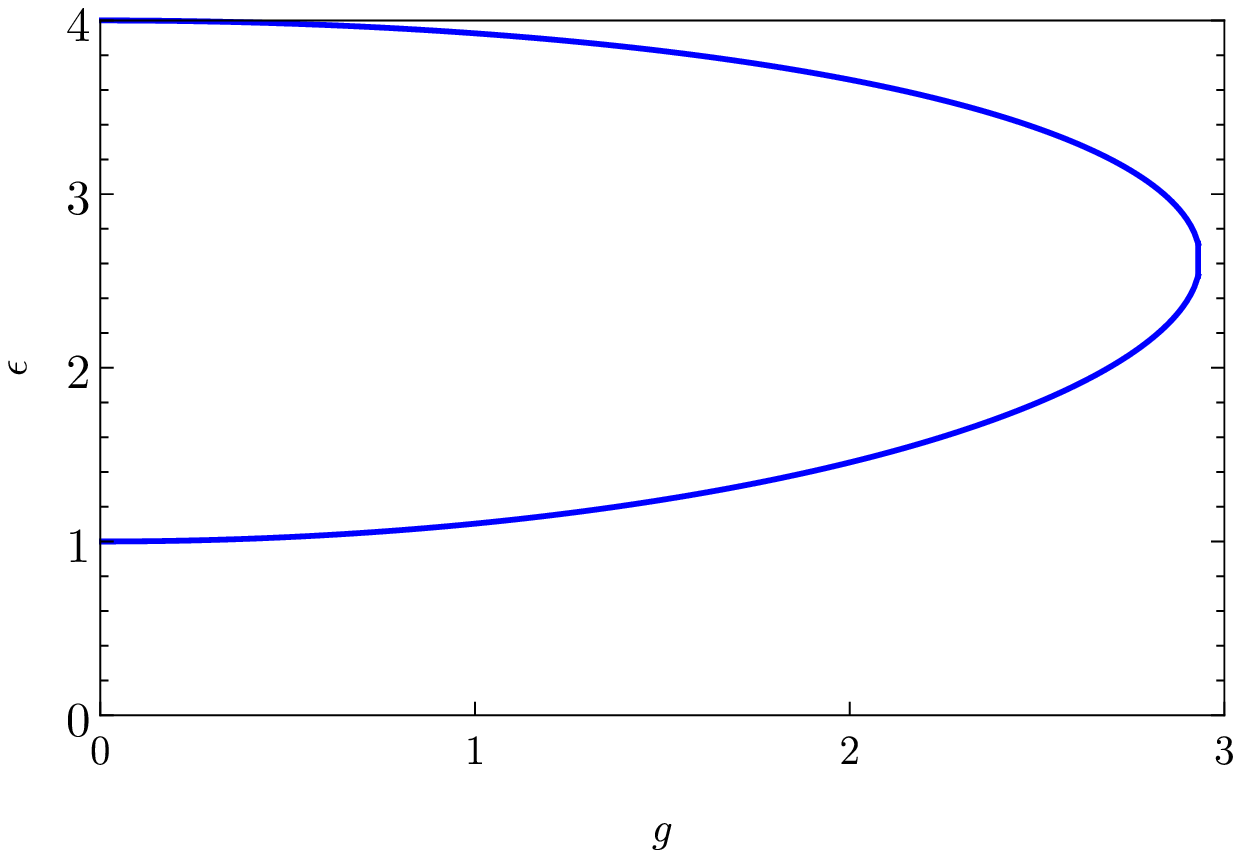}
\par
\includegraphics[width=9cm]{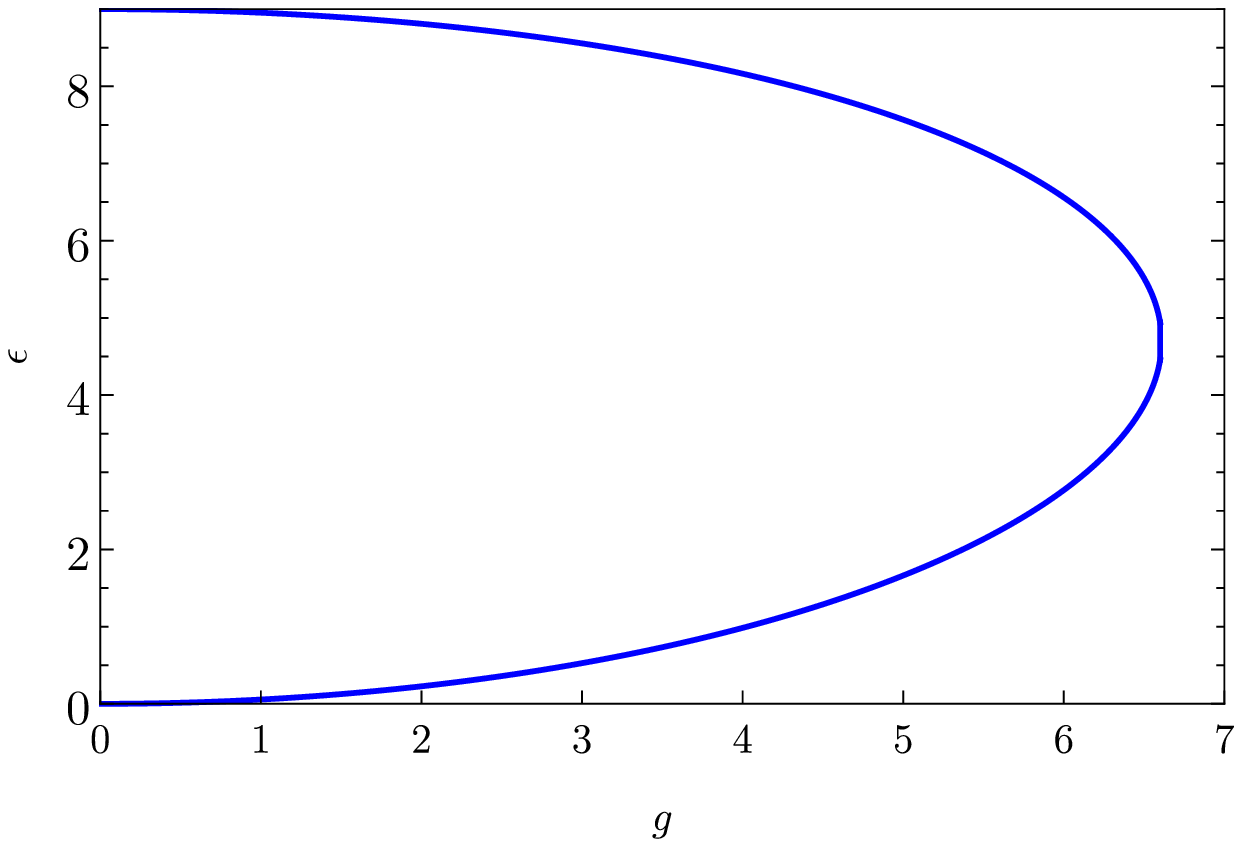}
\end{center}
\caption{First two eigenvalues $\epsilon(ig)$ of symmetry $E$ (upper panel)
and $A$ (lower panel)}
\label{fig:EE_EA}
\end{figure}

\begin{figure}[]
\begin{center}
\includegraphics[width=9cm]{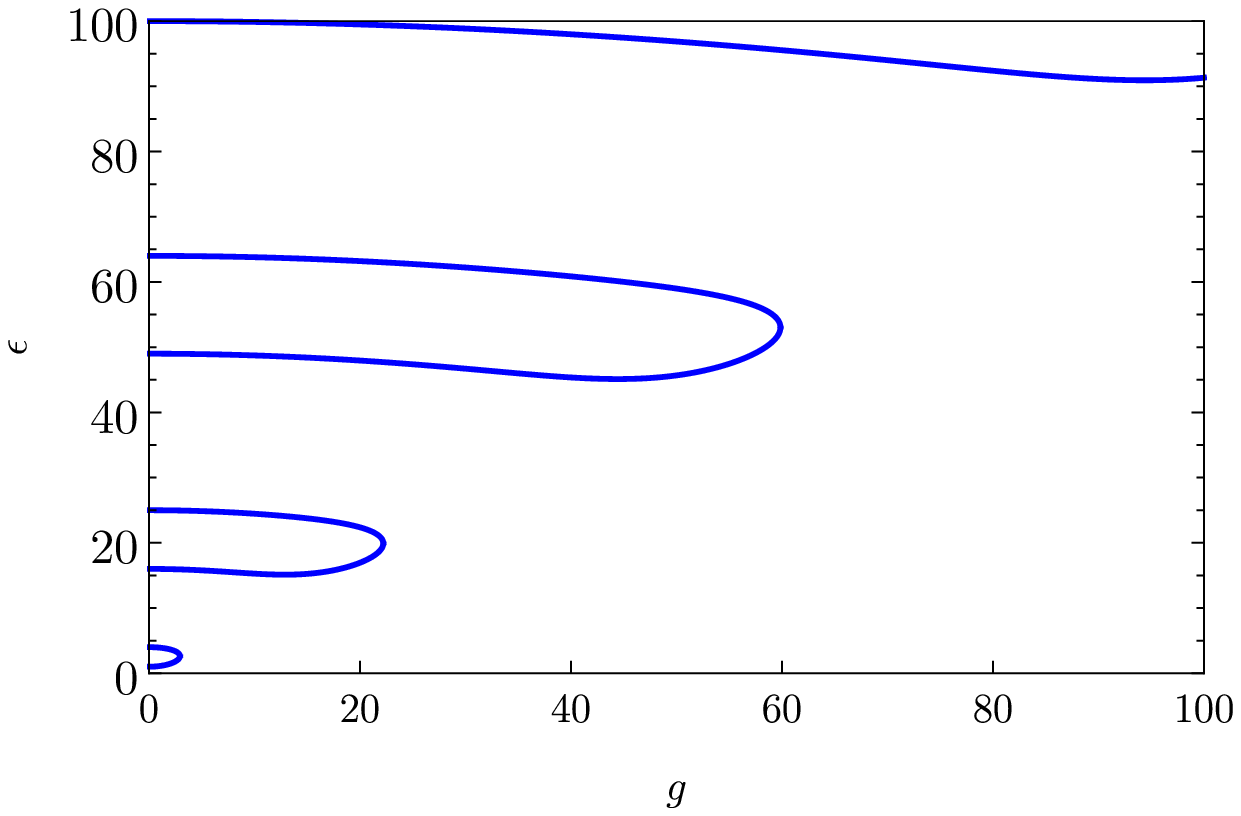}
\par
\includegraphics[width=9cm]{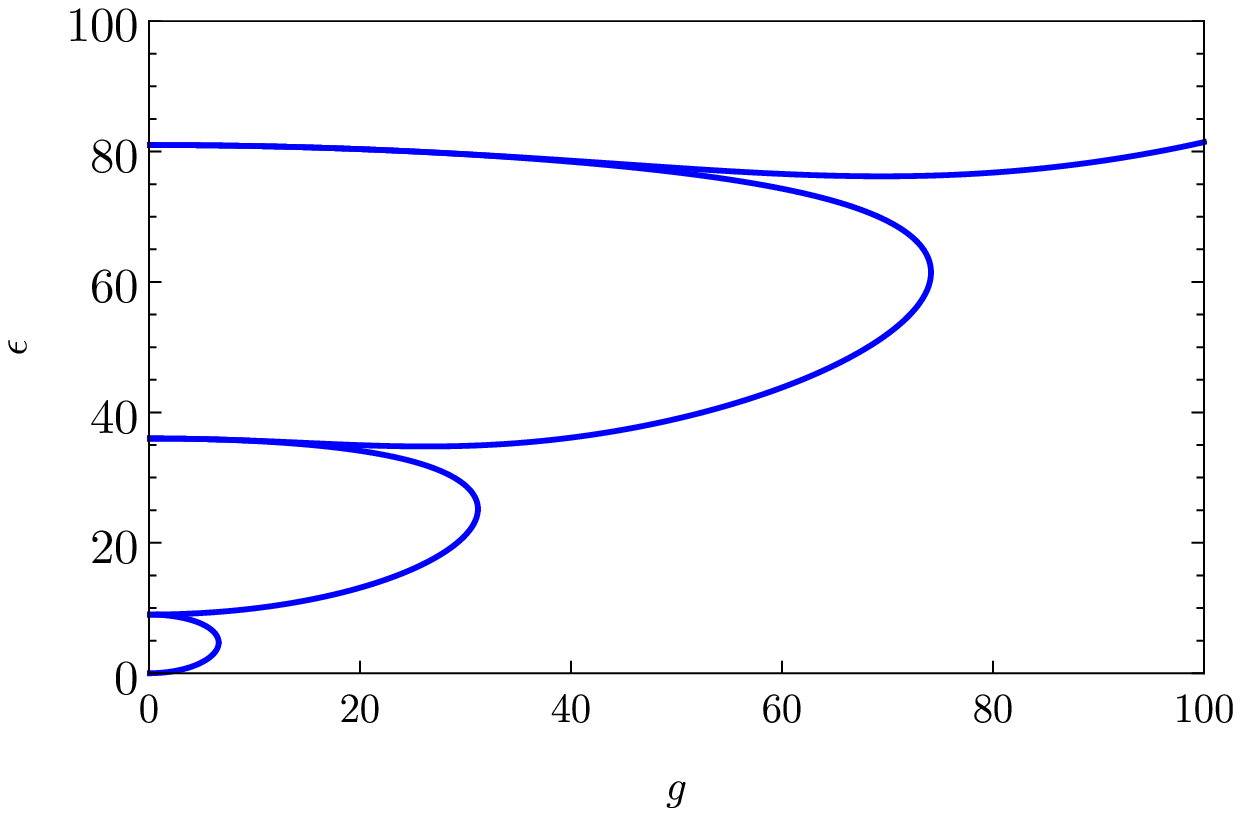}
\end{center}
\caption{Lowest eigenvalues $\epsilon(ig)$ of symmetry $E$ (upper panel) and
$A$ (lower panel)}
\label{fig:EES_EAS}
\end{figure}

Since $\left\langle \cos (3\phi )\psi \right| \left. \cos (3\phi )\psi
\right\rangle \leq \left\langle \psi \right| \left. \psi \right\rangle $ for
all $\psi $ the series (\ref{eq:epsilon_lambda_series}) has a finite radius
of convergence\cite{RS78} and $\epsilon _{n}(ig)$ will be real in the region
of analyticity. More precisely, a given eigenvalue $\epsilon (ig)$ is real
for all $|g|<$ $|g_{e}|$ where $g_{e}$ is an exceptional point where two
eigenvalues coalesce as shown in Figure~\ref{fig:EE_EA} for the two lowest
eigenvalues of symmetry $E$ and $A$. For $|g|>|g_{e}|$ the coalescing
eigenvalues become a pair of complex conjugate numbers. There are simple and
efficient numerical methods for the calculation of the exceptional points
for quantum mechanical models similar to this one\cite{F01}; for the first
two $E$ and $A$ states shown in Figure~\ref{fig:EE_EA} we obtained $%
|g_{e1}|=2.9356105095073260590$, $\epsilon
(g_{e1})=2.6226454301444952679$ and
$|g_{e2}|=6.6094587620331389653$, $\epsilon
(g_{e2})=4.6995725311868146666$, respectively.
Figure~\ref{fig:EES_EAS} shows that the exceptional points
increase with the quantum number which leads to
the conclusion that the $\mathcal{ST}$ symmetry is exact for all $%
|g|<|g_{e1}|$.

Some time ago Bender and Kalveks\cite{BK11} and Fern\'{a}ndez and Garcia\cite
{FG14} discussed other space-time-symmetric hindered rotors with somewhat
different symmetries and calculated several exceptional points. In
particular, the latter authors estimated the trend of the location of the
exceptional points in terms of the quantum numbers of the coalescing states.

\section{Conclusions}

\label{sec:conclusions}

We have shown that the use of parity considerably simplifies the calculation
of the eigenvalues with eigenfunctions of symmetry $A$ of the restricted
rigid rotor with $C_{3}$ symmetry. This strategy is particularly useful in
the case of small barriers o large quantum numbers. We are aware that this
situation is not commonly encountered in most physical applications of the
model\cite{CHHP84,PH97,H99,HH85,KS16,KS17} but we think that it is worth
taking into account the difficulties that it may rise.

We have also shown that this simple model exhibits a non-Hermitian $\mathcal{%
ST}$-symmetric counterpart with real eigenvalues for sufficiently small $%
|\lambda |=|g|$ and obtained the exceptional point $g_{e}$ that determines
the phase transition between exact and broken $\mathcal{ST}$ symmetry. In
this way we added another member to the family of similar problems intensely
investigated in the last years\cite{B07} (and references therein).

\begin{thebibliography}{99}
\bibitem{CHHP84}  S. Clough, A. Heidemann, A. H. Horsewill, and M. N. J.
Paley, Coupled tunnelling motion of a pair of methyl groups in lithium
acetate studied by inelastic neutron scattering, Z. Phys. B: Condens Matter
55 (1984) 1-6.

\bibitem{PH97}  M. Prager and A. Heidemann, Rotational tunneling and neutron
spectroscopy: A compilation, Chem. Rev. 87 (1997) 2933-2966.

\bibitem{H99}  A.J. Horsewill, Quantum tunnelling aspects of methyl group
rotation studied by NMR, Prog. Nucl. Mag. Reson. Spect. 35 (1999) 359-389.

\bibitem{HH85}  W. H\"{a}usler and A. H\"{u}ller, Tunneling of coupled
methyl groups, Z. Phys.B: Condens Matter 59 (1985) 177-182.

\bibitem{KS16}  S. Khazaei and D. Sebastiani, Methyl rotor quantum states
and the effect of chemical environment in organic crystals: $\gamma $%
-picoline and toluene, J. Chem. Phys. 145 (2016) 234506.

\bibitem{KS17}  S. Khazaei and D. Sebastiani, Tunneling of coupled methyl
quantum rotors in 4-methylpyridine: Single rotor potential versus coupling
interaction, J. Chem. Phys. 147 (2017) 194303.

\bibitem{Z83}  M. Znojil, Potential $r^{2}+\lambda r^{2}/(1+gr^{2})$ and the
analytic continued fractions, J. Phys. A 16 (1983) 293-301.

\bibitem{Z84}  M. Znojil, Schrodinger equation as recurrences: II. General
solutions and their physical asymptotics, J. Phys. A 17 (1984) 1603-1610.

\bibitem{ZST85}  M Znojil, K Sandler, and M Tater, The anharmonic oscillator
problem: a new algebraic solution, J. Phys. A 18 (1985) 2541-2554.

\bibitem{FOT86}  F. M. Fern\'{a}ndez, J. F. Ogilvie, and R. H. Tipping,
Calculation of eigenvalues through recurrence relations, J. Chem. Phys. 85
(1986) 5850-5854.

\bibitem{M76}  A. Messiah, Quantum Mechanics, Vol. II, North-Holland, New
York, (1976).

\bibitem{B07}  C. M. Bender, Making sense of non-Hermitian Hamiltonians,
Rep. Prog. Phys. 70 (2007) 947-1018.

\bibitem{BK11}  C. M. Bender and R. J. Kalveks, Extending $\mathcal{PT}$
Symmetry from Heisenberg Algebra to E2 Algebra, Int. J. Theor. Phys. 50
(2011) 955-962.

\bibitem{FG14}  F. M. Fern\'{a}ndez and J. Garcia, Critical parameters for
non-hermitian Hamiltonians, Appl. Math. Comput. 247 (2014) 141-151.

\bibitem{C90}  F. A. Cotton, Chemical Applications of Group Theory, Third
(John Wiley \& Sons, New York, (1990).

\bibitem{KC08}  S. Klainman and L. S. Cederbaum, Non-Hermitian Hamiltonians
with space-time symmetry, Phys. Rev. A 78 (2008) 062113.

\bibitem{F15}  F. M. Fern\'{a}ndez, Space-time symmetry as a generalization
of parity-time and partial parity-time symmetry, 2015. arXiv:1507.08850
[quant-ph].

\bibitem{W60}  E. Wigner, Normal Form of Antiunitary Operators, J. Math.
Phys. 1 (1960) 409-413.

\bibitem{RS78}  M. Reed and B. Simon, Methods of Modern Mathematical
Physics. IV: Analysis of Operators, Academic Press, New York, (1978).

\bibitem{F01}  F. M. Fern\'{a}ndez, Introduction to Perturbation Theory in
Quantum Mechanics, CRC Press, Boca Raton, (2001).
\end{thebibliography}
\end{document}